\documentclass[conference]{IEEEtran}
\usepackage{amssymb,amsfonts,amsmath}
\usepackage{dsfont}
\usepackage{balance}
\usepackage{cite}

\IEEEoverridecommandlockouts

\ifCLASSINFOpdf
   \usepackage[pdftex]{graphicx}
   \DeclareGraphicsExtensions{.pdf,.jpeg,.png}
\else
\fi
\usepackage{fixltx2e}

\def\(({\left(}
\def\)){\right)}
\def\[[{\left[}
\def\]]{\right]}

\newtheorem{theorem}{Theorem}
\newtheorem{corollary}{Corollary}
\newcommand{\be}{\begin{equation}}
\newcommand{\ee}{\end{equation}}
\newcommand{\bea}{\begin{eqnarray}}
\newcommand{\eea}{\end{eqnarray}}

\def\ind{{\mathbf{1}}}
\def\B{\rm B}
\newcommand{\EE}{{\mathbf{E}}}
\newcommand{\PP}{{\mathbf{P}}}
\newcommand{\RR}{{\mathbb{R}}}
\newcommand{\BEAS}{\begin{eqnarray*}}
\newcommand{\EEAS}{\end{eqnarray*}}
\newcommand{\BEA}{\begin{eqnarray}}
\newcommand{\EEA}{\end{eqnarray}}
\newcommand{\Poi}{ \mathrm{Poi}}

\begin{document}

\title{Spectral Detection in the Censored Block Model}
\author{\IEEEauthorblockN{Alaa Saade}
\IEEEauthorblockA{Laboratoire de Physique Statistique\\
 \'Ecole Normale Sup\'erieure, 24 Rue Lhomond\\
Paris 75005}
\\
\IEEEauthorblockN{Marc Lelarge}
\IEEEauthorblockA{INRIA and \'Ecole Normale Sup\'erieure \\
Paris, France}

\and

\IEEEauthorblockN{Florent
  Krzakala}
\IEEEauthorblockA{
  Sorbonne Universit\'es, UPMC Univ. Paris 06\\
  Laboratoire de Physique Statistique, CNRS UMR 8550 \\
  \'Ecole Normale Sup\'erieure, 24 Rue Lhomond, Paris\\
  }
\\
  
\IEEEauthorblockN{Lenka Zdeborov\'{a}}
\IEEEauthorblockA{Institut de Physique Th\'{e}orique \\ CEA Saclay and URA 2306, \\ CNRS 91191 Gif-sur-Yvette, France.}
}

\maketitle

\begin{abstract}

   We consider the problem of partially recovering hidden
  binary variables from the observation of (few) censored edge
  weights, a problem with applications in community detection,
  correlation clustering and synchronization. We describe two spectral
  algorithms for this task based on the non-backtracking and the Bethe
  Hessian operators. These algorithms are shown to be asymptotically
  optimal for the partial recovery problem, in that they detect the hidden
  assignment as soon as it is information theoretically possible to do
  so.
\end{abstract}

\IEEEpeerreviewmaketitle
\subsection{Introduction}
In many inference problems, the available data can be represented on a
weighted graph. Given the knowledge of the edge weights, the task is
to infer latent variables carried by the nodes.  Here, we shall
consider the problem of recovering binary node labels from censored
edge measurements
\cite{abbe2014decoding,abbe2013conditional}. Specifically, given an
Erd\H{o}s-R\'enyi random graph $G=(V,E)\in {\cal G}(n,\alpha/n)$ with $n$ nodes carrying
latent variables $\sigma_i\!=\!\pm 1,\ 1\!\leq\!i\!\leq\!n$, we draw the edge labels
$J_{ij}\!=\!\pm1,\ (ij)\in E$ from the following distribution:
\begin{equation}
\label{edgeweights}
P(J_{ij}  \lvert \sigma_i,\sigma_j) \!=\! (1-\epsilon) \ind(J_{ij}=\sigma_i \sigma_j) +
\epsilon \ind(J_{ij}=-\sigma_i \sigma_j)\,,
\end{equation}
where $\epsilon$ is a noise parameter. In the noiseless case
$\epsilon\!=\!0$, we have $\sigma_i\sigma_j\! =\!J_{ij}$ and one can easily
recover the communities in each connected component along a spanning
tree. When $\epsilon\!=\!1/2$, on the other hand, the graph doesn't
contain any information about the latent variables $\sigma_i$, and recovery
is impossible. What happens in between? The problem of {\it exactly}
recovering the latent variables $\sigma_i$ has been studied in
\cite{abbe2014decoding}. It turns out that, asymptotically in the
large $n$ limit, exact recovery is shown to be possible if and only if
\begin{align}
\label{exact}
\alpha>\alpha_{\rm exact}= \frac{2\log n}{(1-2\epsilon)^2}\,,
\end{align}
where $\alpha$ is the average degree of the graph. Note that the
variable of an isolated vertex cannot be recovered so that the average
degree has to grow at least like $\log n$, as in the Coupon
collector's problem, to ensure that the graph is connected.

We consider in this paper the case where the average degree $\alpha$
will remain fixed as $n$ tends to infinity. In this setting, we cannot
ask for exact recovery and we consider here a different question: is
it possible to infer an assignment $\hat{\sigma}_i$ of the latent
variables that is \emph{positively correlated} with the planted
variables $\sigma_i$?  We call positively correlated an assignment
$\hat{\sigma}_i$ such that the following quantity, called
\emph{overlap}, is strictly positive:
\begin{align}
\label{overlap}
2\left[{\rm max}\left(\frac{1}{n}\overset{n}{\underset{i=1}{\sum}}\ind(\hat{\sigma}_i=\sigma_i),\frac{1}{n}\overset{n}{\underset{i=1}{\sum}}\ind(\hat{\sigma}_i=-\sigma_i)\right)-\frac{1}{2}\right]\,.
\end{align}

In the limit $n\rightarrow\infty$, this overlap vanishes for a random guess $\hat{\sigma}_i$, and is equal to unity  if the recovery is exact. 
We will refer to the task of finding a positively correlated
assignment $\hat{\sigma}_i$ as \emph{partial recovery}. This task has
been shown \cite{HLM2012,LMX2013} to be possible only if
\begin{align}
\label{transition}
\alpha>\alpha_{\rm detect}=\frac{1}{(1-2\epsilon)^2}.
\end{align}
To the best of our knowledge, there is no rigorous proof that this
bound is also sufficient.  In \cite{HLM2012}, the same authors also
showed that belief propagation (BP) allows to saturate this
bound. However, there is no rigorous analysis of BP for this problem
and the fact that condition (\ref{transition}) is necessary and
sufficient was left as a conjecture in \cite{HLM2012} and only the
necessary part was proved in \cite{LMX2013}. Moreover, from a
practical point of view, BP requires the knowledge of the noise
parameter $\epsilon$.

In this contribution, we describe two simple spectral algorithms and
we show rigorously that they are optimal, in the sense that they can
perform partial recovery as soon as $\alpha>\alpha_{\rm detect}$.
Additionally, the output of these algorithms is shown numerically to
have an overlap similar to that of BP, without requiring the knowledge
of the noise parameter $\epsilon$.  This closes the gap from
\cite{HLM2012,LMX2013}, where spectral methods are introduced that
succeed only if the connectivity is significantly larger than the
threshold (\ref{transition}). The resulting algorithms are thus fast,
trivial to implement, and asymptotically optimal.

\subsection{Motivation and Related work}
There are various interpretations and models that connect to this
problem such as i) Community detection \cite{abbe2013conditional}: we
try to recover the community membership of the nodes based on noisy
(or censored) observations about their relationship; ii) Correlation
clustering \cite{bansal2004correlation}: we try to cluster the graph
$G$ by minimizing the number of ``disagreeing edges'' ($J_{ij}=-1$) in
each cluster. These examples, and others such as synchronisation, are
discussed in details in \cite{abbe2014decoding}.

The inspiration for the present contribution comes from recent
developments in the problem of detecting communities in the (sparse)
stochastic block model. The threshold for partial recovery in the
stochastic block model was conjectured in \cite{decelle2011asymptotic}
and proved in
\cite{mossel2012stochastic,massoulie2013community,mossel2013proof}. Optimal
spectral methods, based on the same operators as the algorithms
introduced here, were proposed in
\cite{krzakala2013spectral,saade2014spectral}. These operators were in
particular shown to be much better suited to very sparse graphs than
the traditional adjacency or Laplacian operators.

Interestingly, this problem first appeared in statistical
physics. Indeed, the posterior distribution corresponding to
eq. (\ref{edgeweights}) reads, using
$\beta_0=\frac 12 \log\frac{1-\epsilon}{\epsilon}$
\begin{align}
\label{spinglass}
P(\sigma\lvert J)=\frac{e^{\beta_0\underset{(i j)\in E}{\sum}J_{ij}\sigma_i\sigma_j}}{\mathcal{Z}_J}\,.
\end{align}
This is nothing but the spin glass \cite{mezard1987spin} problem where
the couplings $J_{ij}$ are correlated with the "planted'' configuration
$\sigma_i$ \cite{PhysRevLett.102.238701,abbe2013conditional}. Such
problems can also be shown to be equivalent to spin glasses on the
so-called Nishimori line
\cite{LectureKrzakala,nishimori1981internal}. With these notations,
the detection condition (\ref{transition}) corresponds to the
well-known spin glass transition \cite{viana1985phase,guerra2004high}
at $\sqrt{\alpha_{\rm detect}}\tanh {\beta_0}=1$. In this spin glass
context, \cite{zhang2014non} already conjectured that a spectral
algorithm based on the non-backtracking operator (see sec. \ref{the
  non-backtracking operator}) was optimal.

\subsection{Outline and main results}
In section \ref{sec:algorithms}, we describe two spectral algorithms
that achieve the threshold (\ref{transition}). These algorithms are
based on two linear operators: the non-backtracking operator
introduced in \cite{krzakala2013spectral}, and the Bethe Hessian
introduced in \cite{saade2014spectral}. We further illustrate their
properties by showing the results of numerical experiments. In section
\ref{sec:properties}, we list the spectral properties of the
non-backtracking operator that are relevant to the present
context. Finally, we discuss the properties of the Bethe Hessian and
its relation with the non-backtracking operator in section
\ref{sec:BH} and discuss its connection with the Bethe free energy.

\section{Spectral algorithms}
\label{sec:algorithms}

\subsection{The non-backtracking operator}
\label{the non-backtracking operator}

The non-backtracking operator acts on the directed edges
$i\rightarrow j$ of the graph as
  \begin{align}
\label{eq:B} 
    {\rm B}_{i\to j,k\to \ell} = J_{k\ell}\ind(j=k) \ind(i\neq \ell) \, . 
\end{align}
It is therefore represented by a $2m\times2m$ matrix, where $m$ is the
number of edges in the graph. As discussed in
\cite{krzakala2013spectral,zhang2014non} the motivation for using this
operator is that it corresponds to the linear approximation of belief propagation for this problem around the so-called uninformative fixed
point of BP.

Similarly to \cite{krzakala2013spectral}, one can show (see
Sec. \ref{sec:BH} for details) that the eigenvalues of ${\rm B}$ that
are different from $\pm 1$ form the spectrum of the simpler
$2n\times 2n$ matrix
\begin{align}
{\rm B}^{\prime}=\left( \begin{array}{cc}
0 & D-\mathds{1} \\
-\mathds{1} & J
\end{array} \right)\,,
\label{defBprime}
\end{align}
where $\mathds{1}$ is the $n\times n$ identity matrix, $D$ is the
diagonal matrix defined by $D_{ii}=d_i$, where $d_i$
is the degree of node $i$, and $J$ has entries equal to the edge weights $J_{ij}$. Furthermore, if $(\lambda\neq\pm1,v\in\mathbb{R}^{2m})$ is an eigenpair of ${\rm B}$, then $(\lambda,v^{\prime}\in\mathbb{R}^{2n})$ is an eigenpair of ${\rm B^{\prime}}$ if
\begin{align}
\label{vectorsBprime1}
&v^{\prime}_{n+i}=\underset{j\in\partial i}{\sum} v_{j\rightarrow i},\qquad \forall 1\leq i\leq n,\\
\label{vectorsBprime2}
&\lambda v^\prime_i=(d_i-1)v^{\prime}_{n+i}\,,
\end{align}
where $\partial i$ and $d_i$ are the set of neighbors and the degree
of node $i$. We will therefore favor using ${\rm B}^{\prime}$. The
algorithm is then as follows: given a graph with edge weights
$J_{ij}$,

\textbf{Algorithm 1}
\begin{enumerate}
\item build the matrix ${\rm B}^{\prime}$
\item compute its leading eigenvalue $\lambda_1$ (with largest magnitude), and its corresponding eigenvector $v^\prime=\{v^\prime_i\}$.
\item if $\lambda_1\in\mathbb{R}$ and $\lambda_1>\sqrt{\alpha}$, where $\alpha$ is the average degree of the graph, set $\hat{x}_i=\text{sign}(v^\prime_{n+i})$. Otherwise, raise an error.
\end{enumerate}
Theorem 1 ensures that whenever (\ref{transition}) holds, this algorithm outputs an assignment $\hat{x}_i$ that is positively correlated with the planted latent variables $x_i$. 

\subsection{The Bethe Hessian}
Another operator closely related to the non-backtracking operator was
introduced in \cite{saade2014spectral}. This operator, called the
Bethe Hessian, is an $n\times n$ real and symmetric matrix defined as
\begin{align}
{\rm H}=(\alpha-1)\mathds{1}-\sqrt{\alpha}J+D\,,
\end{align}
where $D$ is the diagonal matrix of vertex degrees. Based on this
operator, we propose the following algorithm: given a graph with edge
weights $J_{ij}$,

\textbf{Algorithm 2}
\begin{enumerate}
\item build the Bethe Hessian ${\rm H}$
\item compute its (algebraically) smallest eigenvalue $\lambda$, and its corresponding eigenvector $v$.
\item if $\lambda<0$, set $\hat{x}_i=\text{sign}(v_{i})$. Otherwise, raise an error.
\end{enumerate}
Justifications for this second algorithm, and its relation with the
first one, will be provided in section \ref{sec:BH}.  Compared to the
first algorithm, this second one is based on a smaller, symmetric
matrix, which leads to improved numerical performance and
stability. Additionally, in the case of more general edge weights
$J_{ij}\neq\pm 1$, the reduction of ${\rm B}$ to a smaller matrix
${\rm B^{\prime}} $ fails, and one has to work with a $2m\times 2m$
matrix. The Bethe Hessian, on the other hand, generalizes easily to
arbitrary weights without any loss in scalability
\cite{saade2014spectral}.\par

\subsection{Numerical results}
\begin{figure}[!t]
\includegraphics[scale=0.48]{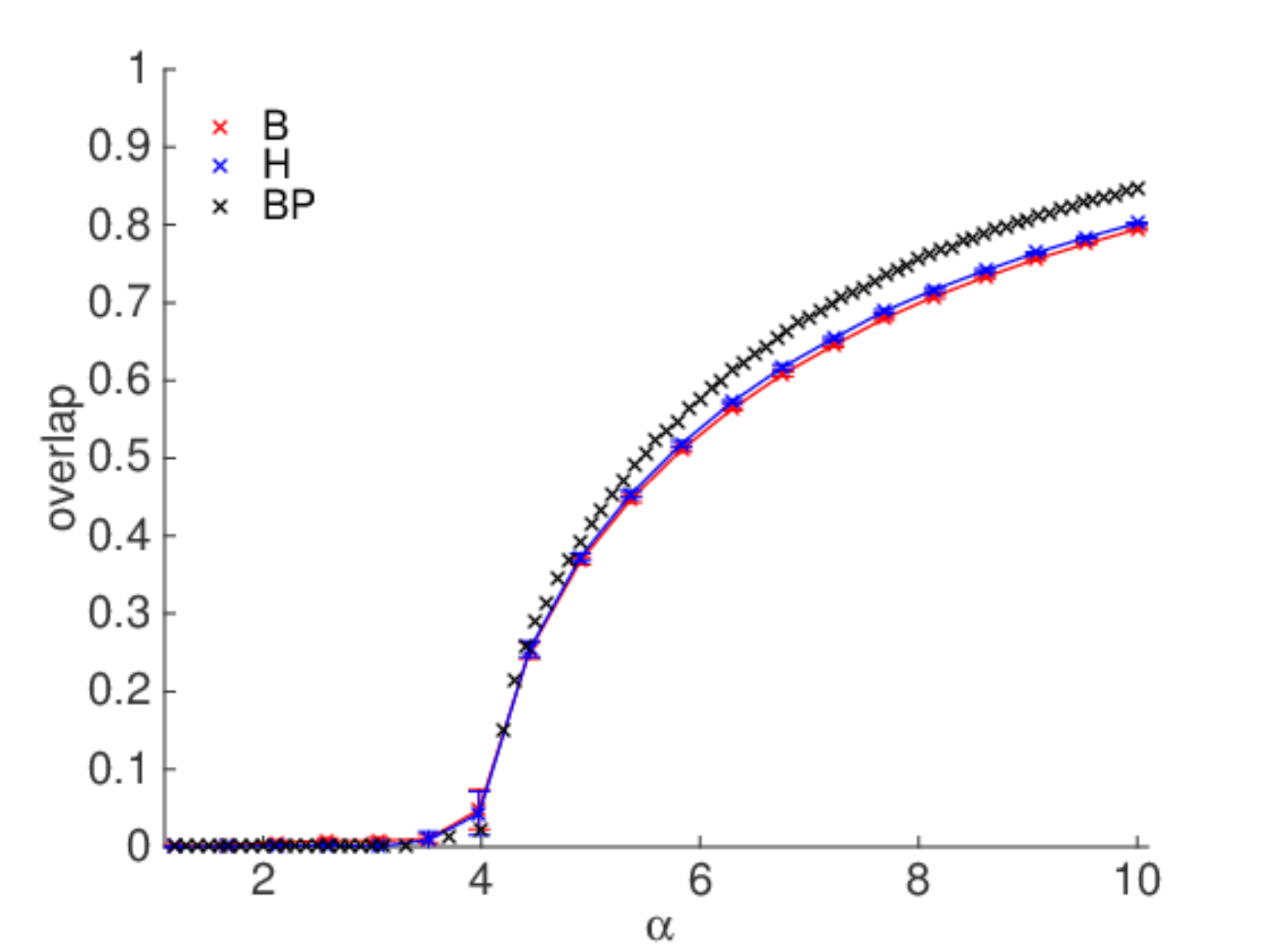}

\caption{Overlap as a function of $\alpha$: comparison between
  algorithm 1 (based on the non-backtracking operator {\rm B}),
  algorithm 2 (based on the Bethe Hessian {\rm H}), and belief
  propagation (BP). The noise parameter $\epsilon$ is fixed to $0.25$
  (corresponding to $\alpha_{\rm detect}=4$), and we vary $\alpha$.
  The overlap for {\rm B} and {\rm H} is averaged over 20 graphs of
  size $n=10^5$. The overlap for BP is estimated asymptotically using
  the standard method of population dynamics (see for instance
  \cite{mezard2009information}), with a population of size $10^4$. All
  three methods output a positively correlated assignment as soon as
  $\alpha>\alpha_{\rm detect}$. Spectral algorithms 1 and 2 have an
  overlap similar to that of BP, with the same phase transition, while
  being simpler and not requiring the knowledge of the parameter
  $\epsilon$. }
\label{fig:overlap}
\end{figure}

Before turning to proofs, we show on figure \ref{fig:overlap} the
numerical performance of our two algorithms, and compare them with the
performance of belief propagation (\cite{pearl1982reverend,HLM2012})
which is believed to be optimal on such locally tree-like graphs in
the sense that it gives, arguably, the Bayes optimal value of the
overlap asymptotically. As shown in section \ref{sec:properties}, both
algorithms 1 and 2 are able to achieve partial recovery as soon as
$\alpha>\alpha_{\rm detect}$, and their overlap is similar to that of
BP, though of course strictly smaller. Note again that BP requires the
knowledge of $\epsilon$ while the two spectral algorithms described
here do not, are trivial to implement, run faster, and avoid the
potential non-convergence problem of belief propagation while
remaining asymptotically optimal in detecting the hidden
assignment. We also observe, empirically, that the overlap given by
the Bethe Hessian seems to be always superior to the one provided by
the non-backtracking operator.

\section{Spectral properties of the non-backtracking operator}
\label{sec:properties}In this section, we state results concerning the spectrum of ${\rm B}$ and show that algorithm 1 outputs an assignment $\hat{\sigma}_i$ that is positively correlated with the planted one, whenever (\ref{transition}) holds. 

As already noticed in previous work for the case of an unweighted random graph \cite{krzakala2013spectral,saade2014spectralDensity}, the superior performance of the non-backtracking operator {\rm B} is due to the particular shape of its spectrum. In the case of the stochastic block model \cite{Holland1983109}, it decomposes into a bulk of uninformative eigenvalues contained in a disk of radius $\sqrt{\alpha}$ in the complex plane, and a few real and informative eigenvalues outside of the disk. This observation was recently proven in \cite{blm15}, in the case of 2 communities. 

The following theorem generalizes this previous result to the present
setting and is the main result of this paper.

\begin{theorem}\label{main:the}
Given an Erd\H{o}s-R\'enyi random graph with average degree $\alpha$, variables assigned to vertices $\sigma_i =\pm 1$ uniformly at
random independently from the graph and where the edges carry weights
sampled from  (\ref{edgeweights}), 
we denote by ${\rm B}$ the non-backtracking operator defined by (\ref{eq:B}).
 and by $\lvert\lambda_1\rvert\geq\lvert\lambda_2\rvert\geq\cdots\geq\lvert\lambda_{2m}\rvert$ the eigenvalues of ${\rm B}$ in order of decreasing magnitude. 
Then, with probability tending to $1$ as $n\to \infty$, we have:
\begin{enumerate}
\item[(i)] if $\alpha<\alpha_{\rm detect}$ then $\lvert\lambda_1\rvert\leq\sqrt{\alpha}+o(1)$.
\item[(ii)] if $\alpha>\alpha_{\rm detect}$, then $\lambda_1\in\mathbb{R},\
  \lambda_1=\alpha(1-2\epsilon)+o(1) > \sqrt{\alpha}$, and $|\lambda_2|\leq\sqrt{\alpha}+o(1)$. 
Additionally, denoting $v$ the eigenvector associated with $\lambda_1$, the following assignment is positively correlated with the planted variables $\sigma_i$:

\BEAS
\hat{\sigma}_i=\text{sign}\left(\underset{j\in\partial i}{\sum} v_{j\rightarrow i}\right)\,.
\EEAS

\end{enumerate}
\end{theorem}

\begin{figure}[!t]
\includegraphics[scale=0.55]{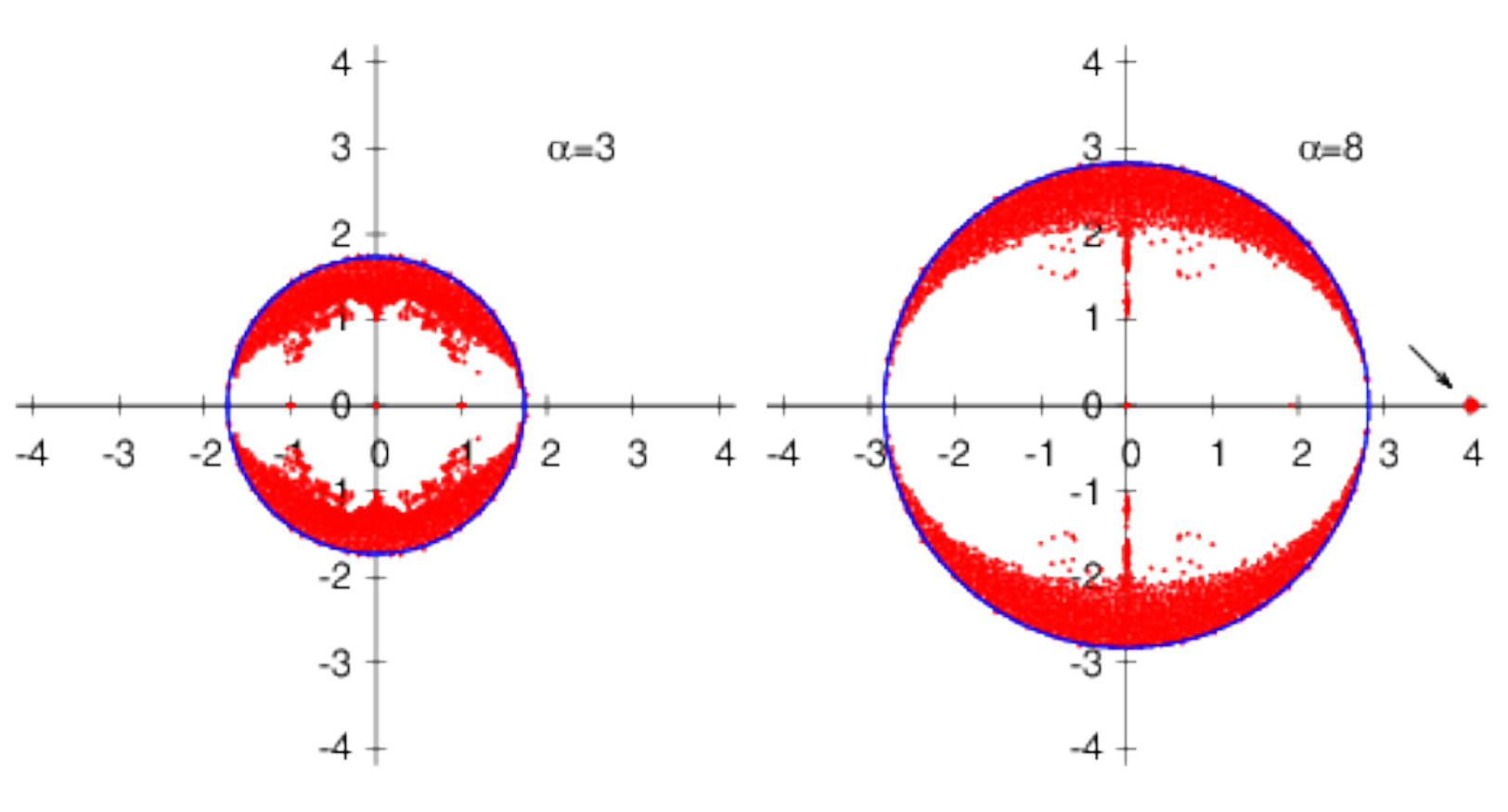}
\caption{Spectrum of the non-backtracking matrix in the complex plane
  for a problem generated with $\epsilon=0.25$, $n=2000$. We used
  $\alpha=3$ (left side) and $\alpha=8$ (right side), to be compared
  with $\alpha_{\rm detect}=4$. Each point represents an eigenvalue. In
  both cases, the bulk of the spectrum is confined in a circle of
  radius $\sqrt{\alpha}$. However, when $\alpha>\alpha_{\rm detect}$,
  a single isolated eigenvalue appears out of the bulk at
  $(1-2\epsilon)\alpha$ (see the arrow on the right plot) and the
  corresponding eigenvector is correlated with the planted
  assignement.}
\label{fig:spectrum}
\end{figure}

This theorem is illustrated on Fig. \ref{fig:spectrum}. 

It is then straightforward to show the following:

\begin{corollary}
  The assignment output by Algo. $1$ is positively correlated with the
  planted variables $\sigma_i$ if and only if
\begin{align}
\alpha> \alpha_{\rm detect}\,.
\end{align}
\end{corollary}

We now give a brief sketch of proof for our Theorem
\ref{main:the}. The proof relies heavily on the techniques developed
in \cite{blm15}.  We try to use notation consistent with \cite{blm15}:
$\vec E$ is the set of oriented edges and for any
$e=u\to v=(u,v) \in \vec E$, we set $e_1=u$, $e_2=v$ and
$e^{-1}=(v,u)$. For a matrice $M$, its transpose is denoted by
$M^*$. We start with a simple observation: if $t$ is the vector in
$\RR^{\vec E}$ defined by $t_e = \sigma_{e_2}$ and $\odot$ is the
Hadamard product, i.e. $(t\odot x)_e = \sigma_{e_2}x_e$, then we have
\BEA
\label{eq:trans}\B x = \lambda x \Leftrightarrow \tilde{B}(t\odot x)= \lambda (t\odot x),
\EEA with $\tilde{\B}$ defined by
$\tilde{\B}_{ef}= \B_{ef}\sigma_{f_1}\sigma_{f_2}$. In particular, $\B$
an $\tilde{\B}$ have the same spectrum and there is a trivial relation
between their eigenvectors. It will be easier to work with $\tilde{\B}$
so to lighten the notation, we will denote (in this section): \BEAS
{\rm B}_{ef} = \ind(e_2=f_1)\ind(e_1\neq f_2)P_f, \EEAS where
$P_f=\sigma_{f_1}J_f\sigma_{f_2}$. Note that the random variables
$P_f$ are now i.i.d. with $\PP(P_f=1)=1-\PP(P_f=-1) =
1-\epsilon$.
With this formulation, the problem is said in statistical physics to
be "on the Nishimori
line"\cite{LectureKrzakala,nishimori1981internal}.

For the case $(1-2\epsilon)^2\alpha <1$, the proof is relatively
easy. Indeed, from \cite{LMX2013}, we know that our setting is
contiguous to the setting with $\epsilon=1/2$. In this case, the random
variable $P_{i,j}$ are centered and a version of the trace method will
allow to upper bound the spectral radius of $B$. Note however, that
one needs to condition on the graph to be $\ell$-tangle-free,
i.e. such that every neighborhood of radius $\ell$ contains at most
one cycle in order to apply the first moment method.

We now consider the case $(1-2\epsilon)^2\alpha \!>\!1$ and 
denote by $P$ the linear mapping on $\RR^{\vec E}$ defined by
$(Px)_e=P_ex_{e^{-1}}$ (i.e. the matrix associated to $P$ is $P_{ef} =
P_e\ind(f=e^{-1})$). Note that $P^*=P$ and since $P_e^2=1$, $P$ is an
involution so that $P$ is an orthogonal matrix. A simple computation
shows that $B^kP = PB^{*k}$, hence $B^kP$ is a symmetric matrix.
This symmetry corresponds to the oriented path symmetry in
\cite{blm15} and will be crucial to our analysis.

We also define $\tilde{\alpha}=(1-2\epsilon)\alpha$ and $\chi\in
\RR^{\vec E}$ with $\chi_e=1$ for all $e\in \vec E$.
The proof strategy is then similar to Section 5 in
\cite{blm15}. Consider a sequence $\ell \sim \kappa
\log_{\tilde{\alpha}}n$ for some small positive $\kappa$. Let
\BEAS
\varphi = \frac{B^\ell \chi}{\|B^\ell \chi\|}, \quad \theta = \|B^\ell
P\varphi\|, \quad \zeta = \frac{B^\ell P\varphi}{\theta}.
\EEAS
If $R= B^\ell - \theta \zeta P\varphi^*$ and we can prove that $\|R\|$
is small in comparison with $\theta$, then we can use a theorem on
perturbation of eigenvalues and eigenvectors adapted from the
Bauer-Fike theorem (see Section 4 in \cite{blm15}) saying that
$B^\ell$ should have an eigenvalue close to $\theta$.

More precisely, for $y\in \RR^{\vec E}$ with $\|y\|=1$, write
$y=sP\varphi +x$ with $x\in (P\varphi)^{\bot}$ and $s\in \RR$. Then,
we find
\BEAS
\|Ry\| =\|B^\ell x+s(B^\ell P\varphi -\theta \zeta)\| \leq \sup_{x:
  \langle x, P\varphi \rangle =0, \|x\|=1}\|B^\ell x\|\,.
\EEAS
This last quantity can be shown to be upper bounded by $(\log
n)^c\alpha^{\ell/2}$ similarly as in Proposition 12 in
\cite{blm15}.
Moreover, we can also show that w.h.p.
\BEA
\label{eq:cond}\langle \zeta, P\varphi\rangle \geq c_0,\quad c_0\tilde{\alpha}^\ell
\leq \theta\leq c_1 \tilde{\alpha}^\ell.
\EEA
These bounds allow to show that $B$ has an eigenvalue $|\lambda_1-
\tilde{\alpha}|=O(1/\ell)$ and that $|\lambda_2|\leq
\sqrt{\alpha}+o(1)$.

Note that $\theta = \frac{\|B^\ell B^{*\ell} P \chi \|}{\|B^\ell
  \chi\|}$, so that we need to compute quantities of the type
$\|B^\ell\chi\|$.
We now explain the main ideas to compute these quantities. First note
that, $(B^\ell\chi)_e$ depends only on the ball of radius $\ell$
around the edge $e$. For $\ell$ not too large, this neighborhood can
be coupled with a Galton-Watson branching process with offspring
distribution $\Poi(\alpha)$. It is then natural to consider this
Poisson Galton-Watson branching process with i.i.d. weights
$P_{u,v}\in\{\pm 1\}$ on its edges with mean $1-2\epsilon$.
For $u$ in the tree, we denote by $|u|$ its generation and by $Y(u)
=\prod_{s=1}^t P_{\gamma_s,\gamma_{s+1}}$ where $\gamma
=(\gamma_1,\dots, \gamma_t)$ is the unique path between the root
$o=\gamma_1$ and $u=\gamma_t$. Then $(B^\ell\chi)_e$ is well
approximated by:
\BEAS
Z_\ell = \sum_{|u|=\ell} Y(u).
\EEAS

It is easy to see that $X_t = \frac{Z_t}{\tilde{\alpha}^t}$ is a
martingale (with respect to the natural filtration) with zero
mean. Moreover we have
\BEAS
\EE\left[ Z_t^2 \right] &=& \EE\left[ \sum_{u,v: |u|=|v|=t}
  Y(u)Y(v)\right]\\
&=& \sum_{i=0}^t \alpha^{t-i} (1-2\epsilon)^{2i}\alpha^{2i}
= O\left( \tilde{\alpha}^{2t}\right),
\EEAS
where the last equality is valid only if $(1-2\epsilon)^2\alpha >1$.
So in this case, we have $\EE\left[X^2_t\right]=O(1)$ and the
martingale $X_t$ converges a.s. and in $L^2$ to a limiting random
variable $X(\infty)$ with mean one.
Following the argument as in \cite{blm15}, this reasoning leads to
(\ref{eq:cond}).

We now consider the eigenvector associated with $\lambda_1$. It
follows from Bauer-Fike theorem (see Section 4 in \cite{blm15}) that
the eigenvector $x$ associated to $\lambda_1$ is asymptotically
aligned with $\frac{B^\ell B^{*\ell} P \chi}{\|B^\ell B^{*\ell} P \chi
  \|}$.
Thanks to the coupling with the branching process, we can prove that
$\|B^\ell B^{*\ell} P \chi \| \approx \tilde{\alpha}^{2\ell}$ and
moreover, we have for $e\in \vec E$,
\BEA
\frac{(B^\ell B^{*\ell} P \chi)_e}{\tilde{\alpha}^{2\ell}} \approx
\frac{\tilde{\alpha}}{\alpha(1-2\epsilon)^2-1}X(\infty),
\label{eq:eigapp}
\EEA
where $X(\infty)$ is the limit of the martingale defined above and has
mean one. We can now translate this result to the eigenvector of the
original non-backtracking operator thanks to (\ref{eq:trans}):
$v_e=\sigma_{e_2}x_e$ where $x_e$ is approximated by
(\ref{eq:eigapp}).
In particular, we see that $\sum_{e,e_2=v}v_e$ is correlated with $\sigma_{v}$.

\section{From the non-backtracking operator to the Bethe Hessian}
\label{sec:BH}
In this section, we relate the spectra of ${\rm H}$, ${\rm B}$ and
${\rm B'}$ by generalizing some properties discussed in
\cite{krzakala2013spectral,saade2014spectral}.
$(\lambda\neq\pm1,v\in\mathbb{R}^{2m})$ being an eigenpair of
${\rm B}$, we define
 \be v_{i}=\underset{j\in\partial i}{\sum}
v_{j\rightarrow i},\qquad \forall 1\leq i\leq n\,. \ee
Since
$\lambda v_{i\to j} = \sum_{k \in \partial i \backslash j} J_{ki} v_{k \to i} $
it follows that $\lambda v_{i\to j}=v_i - J_{ij} v_{j \to i}$. Closing
the equation on the single site elements $v_i$ thus leads to 
\be
\label{key-eq}
v_i
\(( 1 + \sum_{k \in \partial i} \frac {J_{ij}^2}{\lambda-J_{ij}^2} \))
- \lambda \sum_{k \in \partial i} \frac {J_{ij}}{\lambda-J_{ij}^2} v_k
=0 \,.  \ee
For convenience, we now define the matrix:
\begin{align}
{\rm H(X)}=(X^2-1)\mathds{1}-XJ+D
\end{align}
Note in particular that the Bethe Hessian reads
${\rm H}={\rm H}(\sqrt{\alpha})$. Given that the values of $J_{ij}$
are $\pm1$, all eigenvalues of ${\rm B}$ different from $\pm1$ thus must
satisfies the following generalization of the Ihara-Bass
formula\cite{bass1992ihara} :
\begin{align}
\label{ihara-bass}
\det\left[ (\lambda^2-1)\mathds{1}-\lambda J+D \right]= \det {\rm H}(\lambda)=0\,.
\end{align}
To solve (\ref{key-eq}) one needs to find an eigenvector $\bf {v}$ of
${\rm H}(\lambda)$ with a zero eigenvalue. This is a quadratic eigenproblem, which can be turned into a linear one by introducing the matrix ${\rm B^\prime}$ of Algo. 1. Indeed, if
$\lambda\in\mathbb{R}$ is an eigenvalue of ${\rm B^\prime}$ with
eigenvector $v^\prime$, then it follows that
$v:=\{v^\prime_i\}_{n+1\leq i\leq 2n}$ is an eigenvector of
${\rm H}(\lambda)$ with eigenvalue $0$, so that $\lambda$ is an
eigenvalue of ${\rm B}$ as well (at least if  $\lambda \neq \pm 1$), justifying eq. (\ref{vectorsBprime1},\ref{vectorsBprime2}).
Note that since we are
interested in values of $\lambda>1$ (since $\lambda>\alpha$ and we need
$\alpha>1$ from (\ref{transition})), the limitation of looking at
$\lambda \neq \pm1$ is irrelevant.

Finally, following \cite{saade2014spectral}, we can relate the
spectra of ${\rm B}$ and ${\rm H}$ by the following argument. For $X$
large enough, ${\rm H}(X)$ is positive definite. Then as $X$
decreases, ${\rm H}(X)$ will gain a new negative eigenvalue whenever
$X$ becomes equal to an eigenvalue of ${\rm B}$. This justifies the
following corollary: \begin{corollary} if the conditions of
Theorem 1 apply, then ${\rm H}={\rm H}(\sqrt{\alpha})$ has a unique
negative eigenvalue if $\alpha>\alpha_{\rm detect}$, and none
otherwise.\end{corollary}

Strictly speaking, if we denote by $\lambda_1$ the leading eigenvalue
of ${\rm B}$, we have only shown that the eigenvector with eigenvalue
$0$ of ${\rm H}(\lambda_1)$ is positively correlated with the planted
variables if $\alpha>\alpha_{\rm detect}$. However, we observe
numerically (see figure \ref{fig:overlap}) that the eigenvector with
negative eigenvalue of ${\rm H}$ is also positively correlated, and in
fact gives a slightly better overlap. This point will have to be
clarified in future work.

It is worth noting the Bethe Hessian is also related to the belief  propagation algorithm. \cite{yedidia2001bethe} showed that the fixed
points of the BP recursion are stationary points of the so-called
Bethe free energy. Direct optimization of the Bethe free energy has
then been proposed as an alternative to BP
\cite{welling2001belief}. In this context, \cite{saade2014spectral}
showed that the so-called paramagnetic fixed point (corresponding to an 
uninformative assignment) is a local minimum of the Bethe free
energy if and only if ${\rm H}$ is positive definite. Algo. 2 can
therefore be seen as a spectral relaxation of the direct optimization
of the Bethe free energy. In the end, both approaches are indeed deeply
related to BP.

\section{Conclusion}
We have considered the problem of partially recovering binary
variables from the observation of censored edge weights, and described
two optimal spectral algorithms for this task that can provably
perform partial recovery as soon as it is information theoretically
possible to do so. Remarkably, these algorithms do not require the
knowledge of the noise parameter $\epsilon$ and perform almost as well as
belief propagation, which is expected (but not proved) to be Bayes
optimal for this problem. This allows to close the gap from previous
works, both algorithmically, by providing optimal spectral algorithms,
and theoretically, by proving that the transition (\ref{transition})
is a necessary {\it and} sufficient condition for partial recovery.

\balance

\section*{Acknowledgment}
This work has been supported by the ERC under the European Union's FP7
Grant Agreement 307087-SPARCS and by the French Agence Nationale de la
Recherche under reference ANR-11-JS02-005-01 (GAP project).

\bibliography{mybib}

\end{document}